# Small Hemielliptic Dielectric Lens Antenna Analysis in 2-D: Boundary Integral Equations Versus Geometrical and Physical Optics

Artem V. Boriskin, *Member, IEEE,* Gael Godi, Ronan Sauleau, *Senior Member, IEEE,* and Alexander I. Nosich, *Fellow, IEEE*

*Abstract*—We assess the accuracy and relevance of the numerical algorithms based on the principles of Geometrical Optics (GO) and Physical Optics (PO) in the analysis of reduced-size homogeneous dielectric lenses prone to behave as open resonators. As a benchmark solution, we use the Muller boundary integral equations (IEs) discretized with trigonometric Galerkin scheme that has guaranteed and fast convergence as well as controllable accuracy. The lens cross-section is chosen typical for practical applications, namely an extended hemiellipse whose eccentricity satisfies the GO focusing condition. The analysis concerns homogeneous lenses made of rexolite, fused quartz, and silicon with the size varying between 3 and 20 wavelengths in free space. We consider the 2-D case with both *E*- and *H*-polarized plane waves under normal and oblique incidence, and compare characteristics of the near fields.

## I. Introduction

Dielectric hemielliptic lens antennas have been widely used in various applications across a wide range of frequencies (e.g., see [1,2] and references therein). Their design principles are suggested by GO, which states that all the rays that impinge on the lens front interface in parallel to its symmetry axis come together in the rear focal point provided that the eccentricity equals to the inverse value of the refractive index [3]. However, any finite-size lens has a focal spot (not a point) whose size and shape vary with the lens size and material, as well as with the angle of incidence and polarization of the incident wave. Although all lenses are essentially high-frequency devices, electrically small lenses are attractive due to reduced material losses and improved aperture efficiencies [4]. The exact knowledge of the focusing ability and field in focal domain is very important for the design of lens-based sensors.

Manuscript received Jan. 14, 2007. Revised 15 July, 2007.
This work was supported by CNRS via the projects PECO-NEI # 17111 and PICS # 3721, by the Ministry of Education and Science, Ukraine and Ministry of Foreign Affairs, France via the project DNIPRO-82-2005, and by the Conseil Regional de Bretagne via the project CONFOCAL.

A. V. Boriskin and A. I. Nosich are with the Institute of Radiophysics and Electronics NASU, ul. Proskury 12, Kharkiv 61085, Ukraine (e-mail: a_boriskin@yahoo.com).
G. Godi and R. Sauleau are with the "Groupe Antennes et Hyperfréquences", Institut d'Electronique et de Télécommunications de Rennes, Universite de Rennes 1, 35042 Rennes cedex, France.

Among several approaches and techniques developed and applied for the lens antenna analysis, the most widely used are those based on GO and PO [2-13]. The corresponding software is easy to implement, fast, and entails low memory and time expenditures, which is a great merit in comparison to FDTD [14] or IE [15] tools.

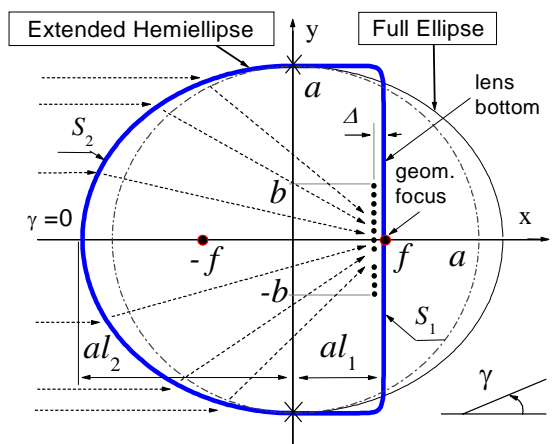

Fig. 1. Geometry and notations of an extended hemielliptic lens. Dotted lines with arrows are for the first-order refracted rays.

In the transmitting mode, GO/PO has been applied successfully for a number of lenses at millimeter (mm) waves. It has been demonstrated that for low-index materials GO/PO design provides a good agreement with measurements, even for electrically small scatterers of complicated shapes [12]. In the receiving mode, no intensive study on the GO/PO relevance for dielectric lens analysis has been reported so far. Although this knowledge is vital for multi-sensor focusing systems, it is still not clear if GO/PO guarantees sufficient accuracy for the near-field analysis of reduced-size lenses, i.e. those whose diameter is smaller than $10 \times \lambda_0$ ($\lambda_0$ is the wavelength in free space), usually considered as an approximate lower limit for the GO/PO methods.

The origins of inaccuracies in GO/PO are well known and relate to neglecting the lens and the field wavefront curvatures and thus a failure to characterize internal resonances of dielectric objects [16]. As a result, the error entering the solution becomes uncontrollable and can significantly affect the near and far field patterns. Although several improvements have been recently introduced to the GO/PO methods [11], their accuracy as to the reduced-size lenses in the receiving mode remains questionable. In contrast to this, the IE-based methods have guaranteed and easily controlled accuracy for arbitrary parameters despite greater programming efforts. This superiority makes them suitable for providing the benchmark numerical solutions, which show that the resonances significantly affect the performance of hemielliptic lenses made of dense materials [17].

The aim of the paper is to quantify the accuracy of GO/PO approach in its conventional formulation in the receiving mode regarding to the near-field analysis, determine its range of validity depending on polarization, size, material and angle of incidence. To this end, we analyze the performance of hemielliptic lens extended with a flat bottom and illuminated with a plane wave (Fig. 1). We concentrate our analysis on the field inside the lens, and try to find out if accounting for multiple internal reflection guarantees sufficient accuracy of the GO/PO algorithm. The final goal is to provide the guidelines for the range of validity of GO/PO for dielectric lens antenna analysis in the receiving mode. The reference solution is obtained by using the numerical algorithm based on the Muller boundary IEs [18,19].

The paper is organized as follows. After outlining the numerical solution methods to be compared in Section II, we present extensive numerical results in Section III. Conclusions are summarized in Section IV.

## II. OUTLINE OF THE SOLUTION METHODS

### A. Geometrical and Physical Optics

A combination of the GO and PO principles is applied to develop a two-step algorithm for the analysis of dielectric lens antennas operating in the receiving mode: (i) a GO ray tracing is used to compute the equivalent electromagnetic currents flowing on the dielectric boundary of the lens and (ii) the near fields are derived from the so-called PO formulation, based on the Huygens-Kirchhoff principle for the 2-D space.

The ray tracing procedure implies a decomposition of the plane wave impinging on the lens interface in terms of elementary ray tubes. Then, the refraction and reflection coefficients are computed with the Fresnel formulas at the intersection points of the ray tubes with the lens boundary. The tubes propagate inside the lens, and the contributions of multiple internal reflections are calculated with the same procedure. Each refraction and reflection of a ray tube at the lens boundary generates equivalent electric and magnetic currents that are integrated with the PO formulas in order to compute the near-field pattern inside the lens. More details can be found in [2].

The main advantages of this approach are four-fold: (i) simplicity of formulation, (ii) very low computational cost, (iii) possibility of determining the influence of each individual internal reflection order on the near-field pattern, and (iv) reasonable accuracy for the large-size lenses.

The main drawback is the evident loss of accuracy for the small-size lenses where the ray tracing is not applicable due to the large curvature of the contour. The approximate nature of the GO/PO technique does not enable one to derive any analytical criterion of its accuracy. This is only possible via comparison with exact numerical results obtained by a method possessing controllable accuracy, such as IE-based ones.

In general, when applying the GO/PO-based algorithm, multiple reflections should be accounted for in the following manner: a ray is assumed reflecting inside the lens until the power contained in the associated tube becomes negligible, namely below 1% of its initial value. The total power stored in the ray tubes after a given number of reflections is the sum of the powers remaining in each ray tube calculated with the Poynting vector formula. The number of reflections necessary to reach this level depends on the lens material and wave polarization. It does not usually exceed 6 for low-index materials such as rexolite and quartz, whereas for high-index materials such as silicon it can be several times larger, especially if a high-$Q$ internal resonance is excited (Fig. 2). As a reasonable compromise, we suggest accounting for not more than the 10 first reflections in order to prevent accumulation of possible errors entering the solution at each reflection.

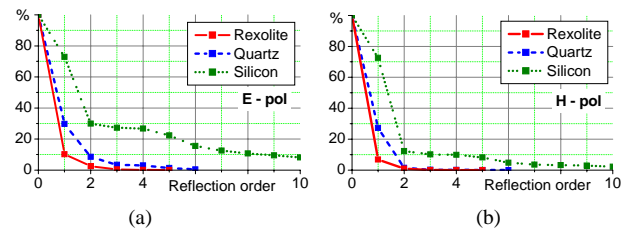

Fig. 2. Percentage of the total power stored in all ray tubes after each reflection in the lenses of Table I. (a) *E*-polarization, (b) *H*-polarization.

TABLE I.
PARAMETERS OF THE LENSES.

| Material | ε | $l_1$ | $l_2$ | $k\Delta$ | $kal_1 - k\Delta$ | $kb$ |
|---|---|---|---|---|---|---|
| Rexolite | 2.53 | 0.81 | 1.29 | 0.40 | 7.60 | 1.97 |
| Quartz | 3.80 | 0.60 | 1.17 | 0.32 | 5.68 | 1.61 |
| Silicon | 11.70 | 0.31 | 1.05 | 0.18 | 2.92 | 0.92 |

### B. The Muller Boundary IEs

The application of the second Green's identity enables one to represent the fields inside and outside a homogeneous dielectric lens in terms of the so-called single- and double-layer potentials, whose density functions are the field and its normal derivative at the lens contour. Here, one has to assume that the contour has continuous normal.

Application of the continuity of the tangential field components across the lens contour leads to a set of the Fredholm second kind IEs known as the Muller Boundary IEs (MBIE) [18]. Then, for a contour of the star-like shape (parameterizable with the polar angle), a discretization scheme based on a Galerkin method with entire-domain angular exponents (also known as trigonometric polynomials) as basis functions can be applied to obtain an infinite-matrix equation with favorable features [19]. To speed up the convergence rate of the algorithm, we also use the concept of analytical regularization [20] in the treatment of integrable singularities in some of the kernel functions and their derivatives. The remaining parts of the matrix and right-hand-part elements are reduced to Fourier-expansion coefficients of the twice-continuous functions that can be economically computed with FFT and DFFT algorithms.

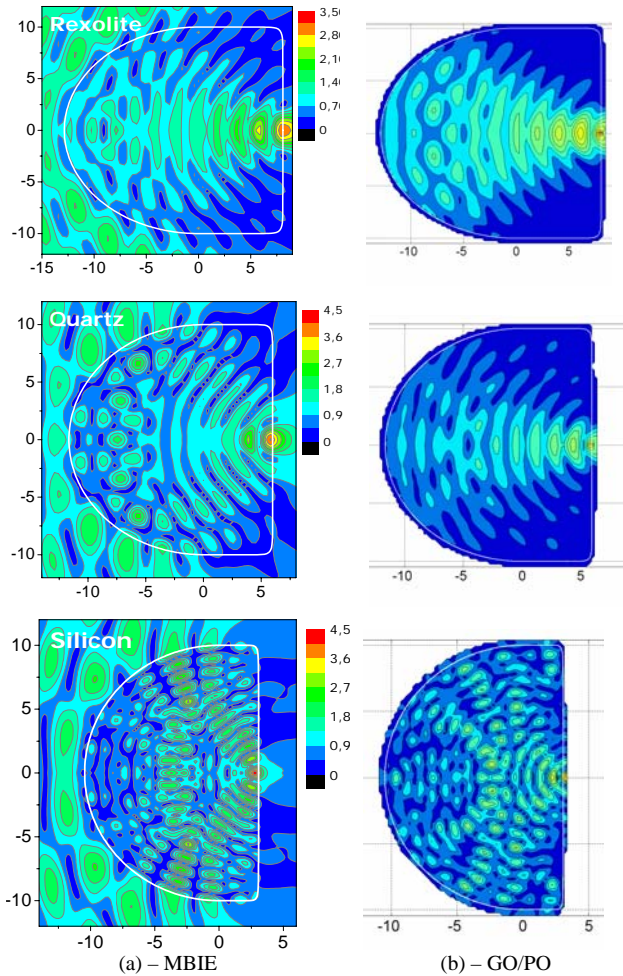

Fig. 3. Field intensity maps inside the hemielliptic lenses computed by (a) MBIE and (b) GO/PO algorithms. The coordinate axes are scaled in $\lambda_0$. The flat bottom of each lens is about $3 \times \lambda_0$ in width ($ka = 10$) and cut through the rear GO focus. The incident field is the *E*-polarized plane wave under symmetrical incidence ($\gamma = 0°$). The lens contour is represented by the white line.

The advantages of this algorithm that make it an efficient and reliable tool for the analysis of dielectric lenses are as follows: (i) controllable accuracy, i.e., possibility to minimize the computational error to a predicted number of digits, determined by DFFT, for an arbitrary set of the problem parameters (including the lens electrical size, shape and material) by solving progressively greater matrices, (ii) stability near and far from the sharp natural resonances, which is hard with conventional numerical approximations, (iii) low memory and time requirements, and (iv) absence of the false "numerical resonances" intrinsic to the IEs methods using field representations in terms of either single- or double-layer potentials [21]. More details of the algorithm properties are available in [19].

### III. NUMERICAL RESULTS

#### A. Small-size lens, normal incidence

In this section, we present the results of the near-field simulations of extended hemielliptic lenses whose eccentricities and flat bottom extensions are chosen to satisfy the GO focusing condition. The selected materials are those most commonly used for the lens antennas in the mm and sub-mm wavelength ranges, namely rexolite ($\varepsilon = 2.53$), quartz ($\varepsilon = 3.80$), and silicon ($\varepsilon = 11.70$). The size of the lens flat bottom varies between $3 \times \lambda_0$ and $4 \times \lambda_0$ that is typical for practical applications [1]. The lenses are excited by a plane wave propagating along the axis of symmetry. Both *E*- (TM) and *H*-polarizations (TE) are considered.

In the simulations, the cross-sectional contour of the extended hemielliptic lens is represented by a twice continuous curve that consists of two parts, $S_1$ and $S_2$, smoothly joined together at the points marked with crosses (Fig. 1). Here, $S_2$ is a half of the ellipse with eccentricity $e = 1/\sqrt{\varepsilon}$ and $S_1$ is a half of so-called "super-ellipse", that is a rectangle with rounded corners [17,19].

It is not so obvious what quantity should be used as a figure of merit for the lens focusing ability. We have introduced a virtual aperture of the length $2b$ ($b = \lambda_e/2$, $\lambda_e$ is the wavelength in dielectric) placed inside the lens perpendicular to the axis of symmetry at the distance $\Delta$ from the lens bottom ($\Delta = \lambda_e/10$). The integral intensity of the field at this aperture can be calculated as follows:

$$I = \int_{-kb}^{kb} |\Psi(kx_a, ky)|^2 dky, \quad x_a = al_1 - \Delta \quad (1)$$

where $\Psi = E_z$ or $H_z$ for *E*- or *H*-polarization, respectively, and characterizes the efficiency of the focusing.

In order to visualize the near-fields inside the lens and assess the accuracy of the GO/PO method, the near-field maps have been plotted for the three lenses (Fig. 3). Note that all near-field characteristics presented in this paper are normalized by the amplitude of the incident plane wave (for the IE results) or by a "best-fit" coefficient (for the GO/PO results). For convenience, the corresponding values of the normalized geometrical parameters of all three lenses are given in Table 1.

As one can see, internal reflections play significant role in the behavior of small-size hemielliptic dielectric lenses. Although accounting for multiple reflections enables one to obtain the near-field maps that visually resemble the exact MBIE ones, GO/PO is not capable of describing the electromagnetic behavior in full manner, especially for the lenses made of the high-index materials and/or near the rear focus of the lens.

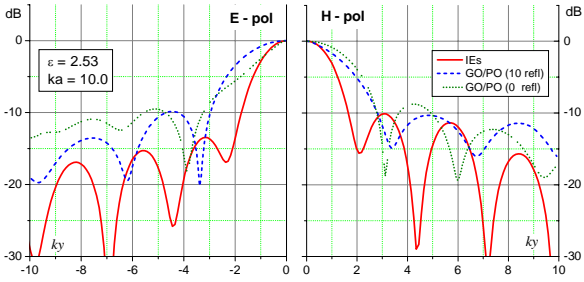

Fig. 4. Normalized field intensity inside the rexolite lens in the plane $x=x_a$ ($\gamma = 0°$). The GO/PO curves are obtained with and w/o accounting for internal reflections.

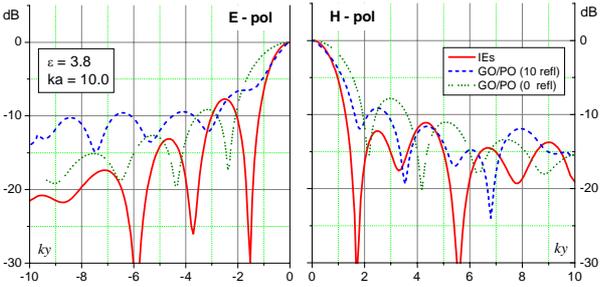

Fig. 5. The same as in Fig. 4, but for the quartz lens.

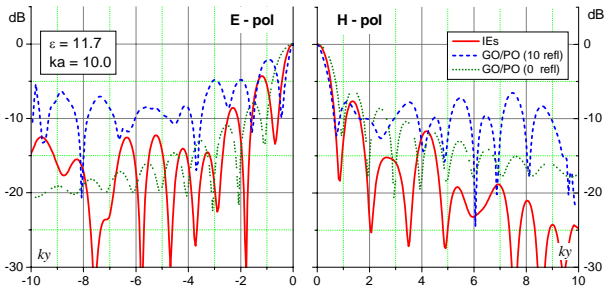

Fig. 6. The same as in Fig. 4, but for the silicon lens.

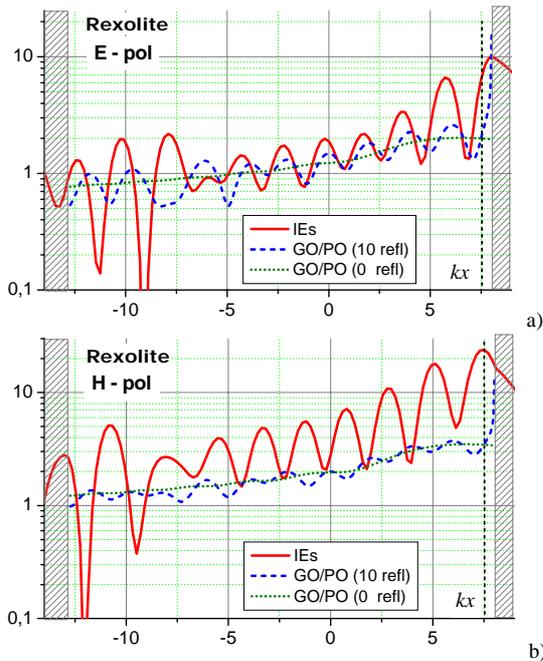

Fig. 7. Intensity of the field inside the rexolite hemielliptic lens along its axis of symmetry. The shadow regions on the left and right are bordering the area inside the lens. The vertical dashed line indicates the location of the virtual aperture inside the lens. The GO/PO curves are obtained with and w/o accounting for internal reflections.

To study fine details of the field in the focal domain, we plotted the normalized field intensities $|\Psi|^2$ along the lens flat bottoms (Figs. 4-6). Here, the left and the right sides of the plots correspond to the *E*- and *H*-polarizations, respectively.

As one can see, the difference between the curves computed by the GO/PO and the MBIE algorithms is quite significant even for rexolite. To explain this we have also plotted the field intensity along the axis of symmetry of the lens (Figs. 7-9).

Again, it is well seen that that GO/PO fails to provide realistic information about the focal spot size, location and peak field value. Although accounting for multiple reflections improves the performance of the algorithm outside of the focal domain, this cannot compensate for the inaccuracies near to either of the ellipse foci.

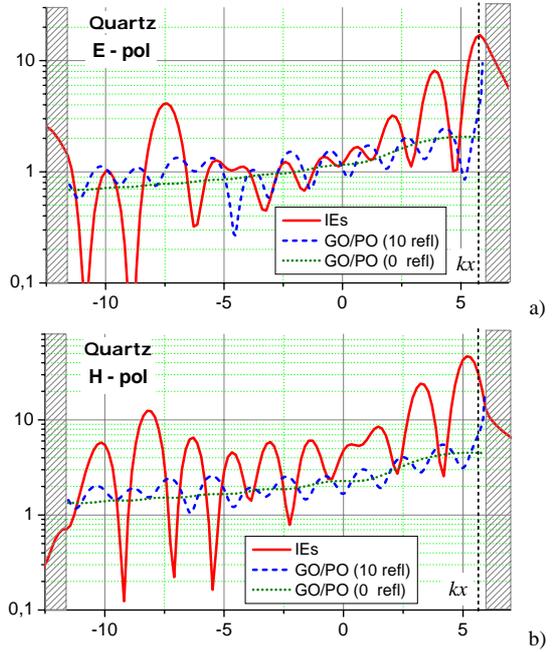

Fig. 8. The same as in Fig. 7 but for the quartz lens.

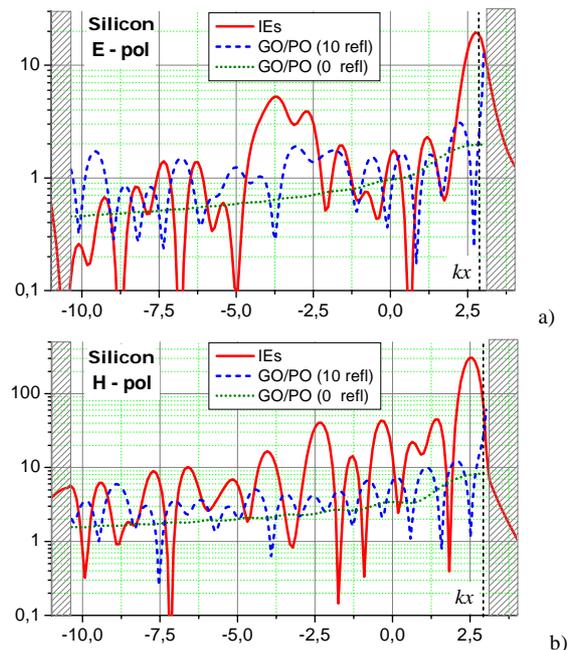

Fig. 9. The same as in Fig. 7, but for the silicon lens.

*B. Medium-size lens, normal incidence*

The plots of the integral intensity in the virtual aperture defined by Eq. (1) versus the normalized frequency *ka* for the above discussed lenses are presented in Figs. 10-12. As one can see, the curves obtained with GO/PO and MBIE are significantly different both in absolute values and in shape. The former is due to the inaccurate description of the focal spot size in GO/PO that results in the lower level of intensity at the apertures (see Figs. 7-9). The latter shows inaccuracies in the description of internal resonances that entails errors in the field intensity in the focal region.

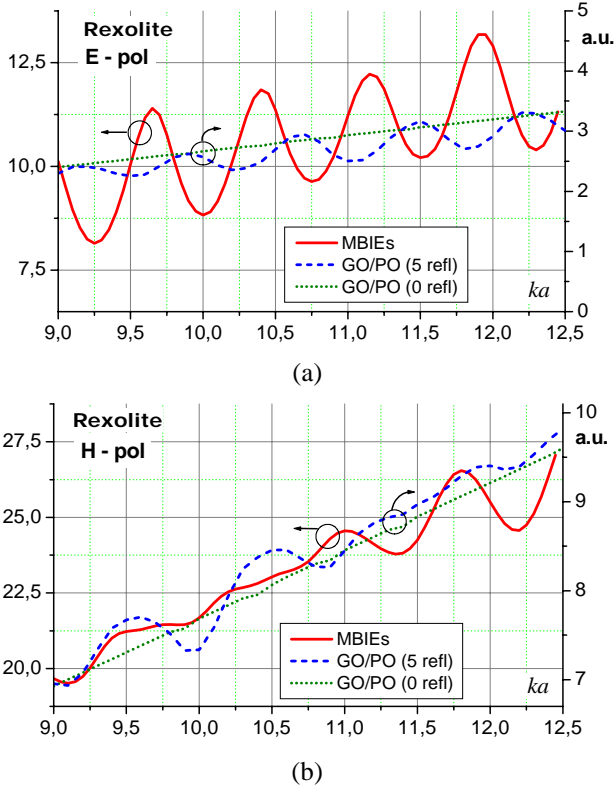

Fig. 10. Normalized integral intensity in the virtual aperture located inside rexolite hemielliptic lens ($x = x_a$, $\gamma = 0°$). The curves for GO/PO are obtained with and w/o accounting for the multiple internal reflections. The lens flat-side size varies between $2.8 \times \lambda_0$ and $4 \times \lambda_0$.

The periodic oscillations well observed in Figs. 10-12 for the curves obtained by MBIE and by GO/PO-plus-multiple-reflections highlight strong involvement of the internal resonances in the behavior of medium-size lenses even for a low permittivity material such as rexolite. It is also well seen that for the silicon lens (Fig. 12), the focusing ability is even stronger affected by the higher-*Q* resonances not found in rexolite and quartz. These resonances are more pronounced in the *E*-polarization case, than in the *H*-case.

The near field for one of such resonances, computed with MBIE algorithm, is presented in Fig. 13.

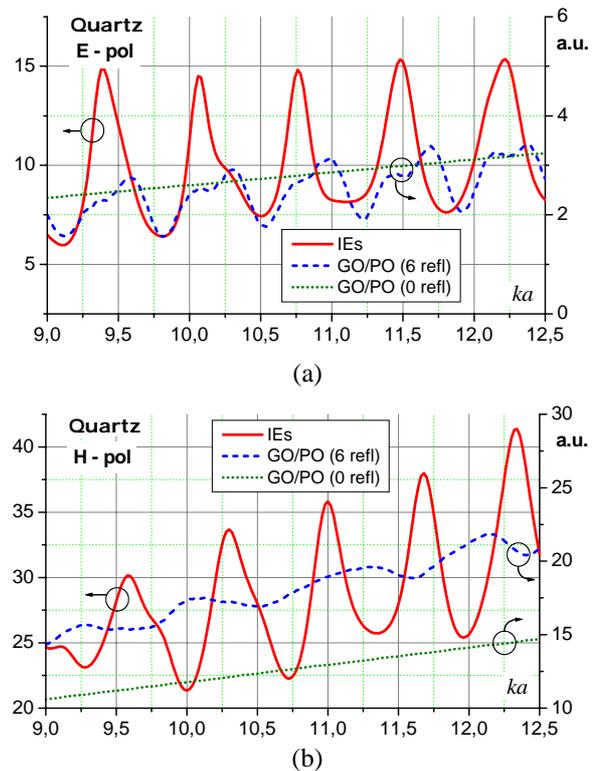

Fig. 11. The same as in Fig. 10, but for quartz lenses.

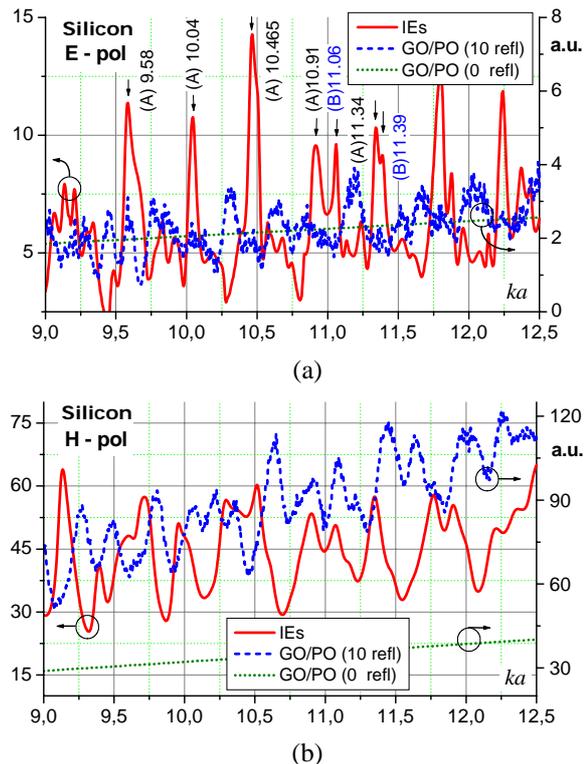

Fig. 12. The same as in Fig. 10, but for silicon lenses.

Comparison of this pattern with the one presented in Fig. 3 (left-bottom one) shows that these resonances are of the same family and differ only by the number of field spots along a certain curved triangle (see also [17]). The distance between the two neighboring E-type resonances in Fig. 12 is $\delta_A \approx 0.42$ and keeps this value for all resonances of the same family indicated by the letter A. The other resonances observed in Fig. 12 are indicated as B-type and belong to another family having different periodicity of spikes ($\delta_B \approx 0.33$). The near-field map for one of the B-type resonances is given in Fig. 14. It is difficult to classify this resonance based only on its map without solving the corresponding eigen-value problem. Nevertheless, as one can see, the resonances of both types dramatically affect the lens performance and cannot be accurately predicted by using GO/PO.

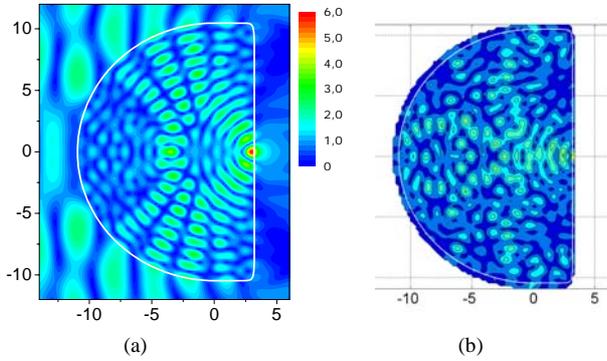

Fig. 13. Near-field maps of the silicon lens with $ka$=10.46 illuminated by a normally incident plane *E*-wave: (a) MBIE, (b) GO/PO.

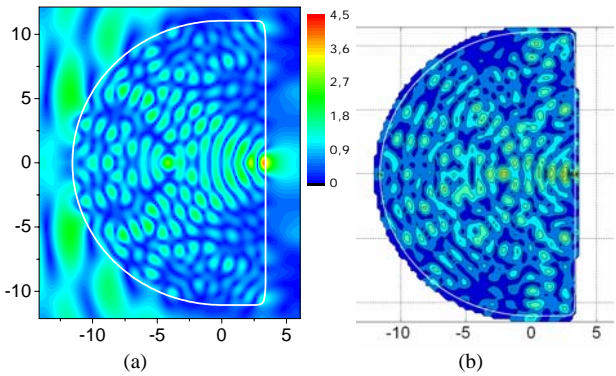

Fig. 14. The same as in Fig. 13, but for $ka$ =11.06.

In order to validate the GO/PO performance for the lenses of larger electrical sizes, we have plotted the integral intensity in the virtual aperture for a rexolite lens with the flat bottom size up to $20 \times \lambda_0$ (Fig. 15). As one can see, for the larger lenses, the resonances start playing less important role due to smaller values of resonance spikes and integral intensity in the aperture, which grow proportionally to the lens size. A lower level of intensity and larger amplitudes of the resonant spikes observed for the *E*-polarization can be explained by the fact that the Fresnel formulas for the reflection and transmission coefficients at the air/dielectric interface suggest that it is less penetrable for the *E*-polarized waves than for the *H*-ones.

*C. Oblique incidence*

Analyzing the curves for the normalized field intensity along the lens flat bottom for oblique incidence (Figs. 16-18), one can see that GO/PO, in most cases, is capable of predicting the location of the focal spot center.
However, it does not provide reliable information about the focal domain size and shape and on the side-spots level. Accounting for the multiple internal reflections improves the accuracy of GO/PO algorithm as to the main focal spot size and shape description however it does not improve the accuracy of side-spot description; moreover, in some cases it works in opposite way (Fig. 16-b).

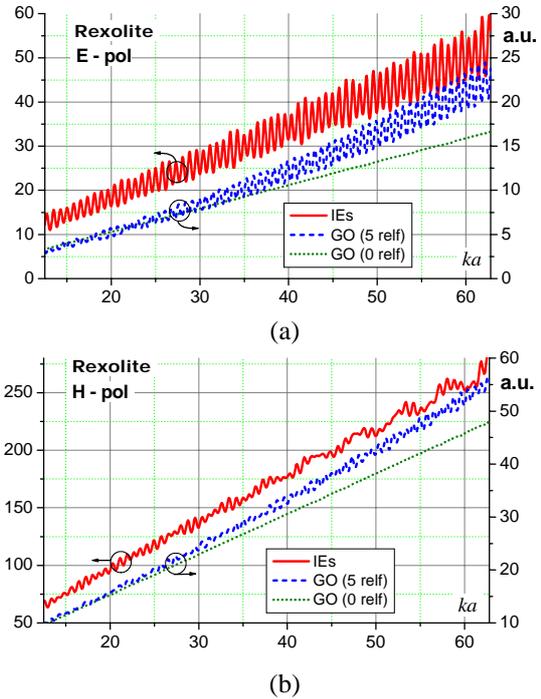

Fig. 15. Normalized integral intensity in the virtual aperture inside the rexolite lens as shown in Fig. 1 vs. normalized frequency. GO/PO curves are obtained with and w/o accounting for the internal reflections.

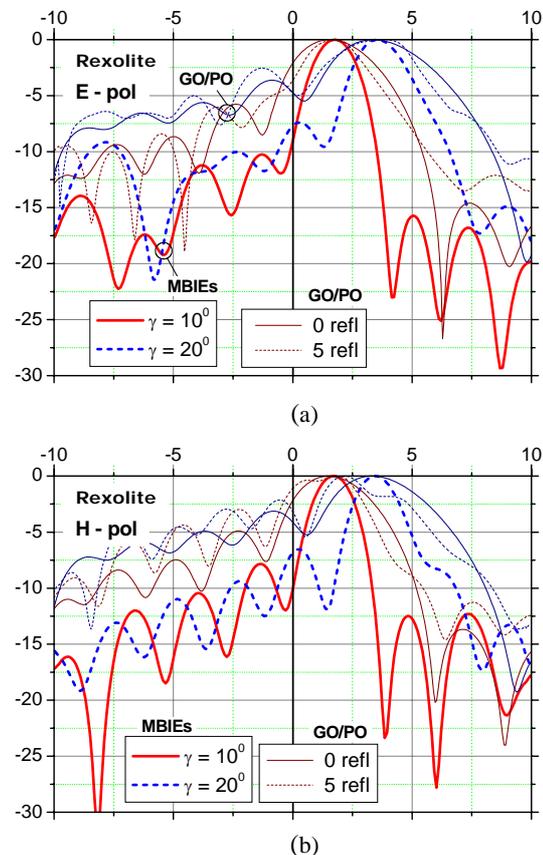

Fig. 16. Normalized field intensity function inside the rexolite lens in the plane coinciding with the aperture ($x=x_a$). The GO/PO curves are obtained with and w/o accounting for internal reflections.

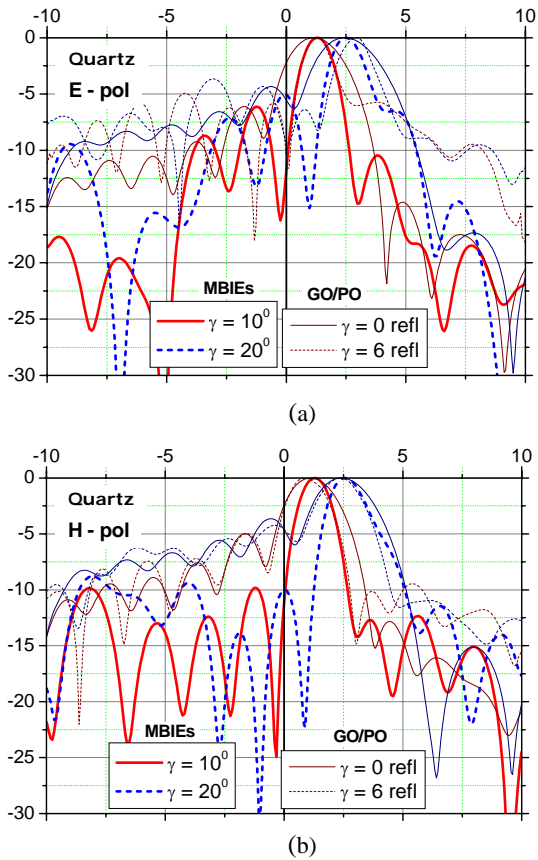

Fig. 17. The same as in Fig. 16, but for quartz lenses.

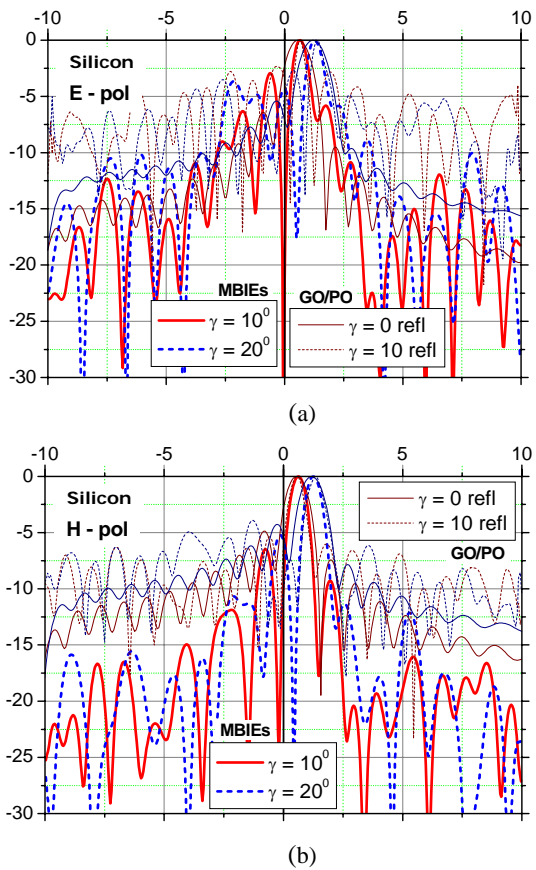

Fig. 18. The same as in Fig. 16, but for silicon lenses.

## IV. CONCLUSIONS

Two numerical algorithms based on the GO/PO and MBIE techniques have been implemented and applied to the in-depth analysis of the near fields inside the small-to-medium size hemielliptic homogeneous dielectric lenses illuminated by the *E*- and *H*-polarized plane waves. The algorithm based on MBIEs provides controllable accuracy for arbitrary set of lens parameters and has been used as a reference solution. We have paid special attention to the resonance phenomena in hemielliptic lenses. Different types of internal resonances excited within the lens have been demonstrated and their influence on the lens focusing ability has been studied for the rexolite, quartz and silicon lenses.

Our analysis has revealed that inaccuracies in the GO/PO algorithm as to the internal resonance simulation cannot be easily compensated by the accounting for the multiple reflections. These inaccuracies relate to focal domain size, shape and the level of the field intensity along the lens flat bottom. On the basis of presented results, we do not recommend to use the GO/PO technique for the near-field analysis of small-to-medium size dielectric lenses in the receiving mode.

Here we would like to note that these results and conclusions are not directly applicable for the transmitting mode. In that mode, the amplitudes of equivalent currents induced in the aperture by the reflected rays are usually much smaller than the amplitudes of the initial currents as they are proportional to the power stored in the reflected rays. This value, for low-density materials, does not exceed several percent [6] and thus cannot significantly affect the lens far-field characteristics, e.g. their radiation patterns. Nevertheless, one has to be aware that for the higher-index materials such as silicon, high-*Q* resonances can spoil the accuracy of analysis in the same manner as in the receiving mode. The validation of the GO/PO accuracy for the transmitting mode will be studied in a forthcoming paper.

### ACKNOWLEDGMENT

The authors are grateful to Dr. S. V. Boriskina for many valuable discussions and also thank CNRS/IDRIS for the access to their computing platforms.

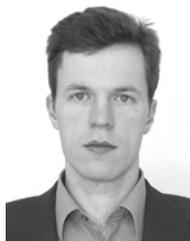

**Artem V. Boriskin** (S'99-M'04) was born in Kharkov, Ukraine, in 1977. He received the M.S. degree in radio-physics from the Kharkov National University in 1999 and Ph.D. degree from the Institute of Radiophysics and Electronics, National Academy of Sciences of Ukraine (IRE NASU) in 2004, respectively.

Since 2003 he has been on research staff of the Department of Computational Electromagnetics, IRE NASU. Since 2004 he has held a CNRS visiting postdoctoral position at the Institute of Electronics and Telecommunications of Rennes (IETR), University of Rennes 1, France. His research interests are in development of numerical algorithms for analysis and optimization of arbitrary-shaped dielectric scatterers with application to dielectric antennas and lenses.

Dr. Boriskin received the IEEE MTT-S Graduate Student Fellowship Award in 2000, the student fellowship of NATO and Turkish Council of Science and Technology in 2001, the 1st Prize of the European Microwave Association and MSMW-04 Symposium in 2004, and the Young Scientist Research Grant of the President of Ukraine in 2006.

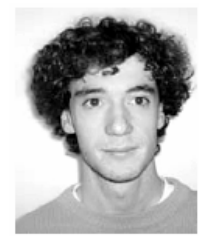

**Gaël Godi** was born in Charleville-Mézières, France in 1980. He received the M.Sc. degree in electronics from the Institut National des Sciences Appliquées, Rennes, France, in 2003.

Since 2003, he has been working toward the Ph.D. degree in signal processing and telecommunications at the IETR, University of Rennes 1, Rennes, France. His main fields of interest include the analysis and optimization of lens antennas for mm-wave applications.

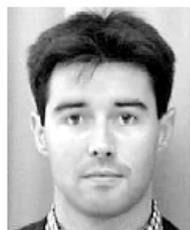

**Ronan Sauleau** (M'04, SM'06) received the Electronic Engineering and Radiocommunications degree and the French DEA degree in electronics from INSA, Rennes, France, in 1995, the aggregation from Ecole Normale Supérieure de Cachan, France, in 1996, and the Doctoral degree in signal processing and telecommunications from IETR, University of Rennes 1, Rennes, in 1999.

Since 2000, he had been an Assistant Professor at the University of Rennes 1. He was elected as an Associate Professor in 2005. His main fields of interest are millimeter wave printed antennas, focusing devices, and periodic structures including electromagnetic bandgap materials and metamaterials.

Dr. Sauleau was the recipient of the 2004 ISAP conference Young Scientist Travel Grant and the first Young Researcher Prize in Brittany, France in 2001 for his work on gain-enhanced Fabry-Perot antennas.

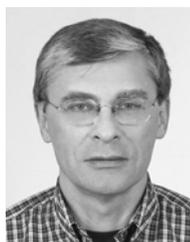

**Alexander I. Nosich** (M'94–SM'95–F'04) was born in Kharkov, Ukraine, in 1953. He received the M.S., Ph.D., and D.Sc. degrees in radio physics from the Kharkov National University in 1975, 1979, and 1990, respectively. Since 1979, he has been with IRE NASU, Kharkov, where he is currently a Professor and Leading Scientist.

Since 1992, he has held a number of guest Fellowships and Professorships in the EU, Japan, Singapore, and Turkey. His research interests include the method of analytical regularization, propagation and scattering of waves in open waveguides, simulation of semiconductor lasers and antennas, and the history of microwaves.

Dr. Nosich was one of the initiators and technical committee chairman of the international conference series on Mathematical Methods in Electromagnetic Theory (MMET) held in the USSR and Ukraine in 1990-2006. In 1995, he organized an IEEE AP-S Chapter in East Ukraine, the first one in the former USSR. From 2001 to 2003, he represented the Ukraine, Poland, and the Baltic States in the European Microwave Association.